\documentstyle[12pt]{article}
\newcommand{\eqnum}[1]{\setcounter{equation}{#1}}

\newcommand{\non}{\nonumber}
\newcommand{\beq}{\begin{equation}}
\newcommand{\eeq}{\end{equation}}
\newcommand{\barr}{\begin{eqnarray}}
\newcommand{\earr}{\end{eqnarray}}

\newcommand{\square}{\kern1pt\vbox{\hrule height 1.2pt\hbox
{\vrule width1.2pt\hskip 3pt \vbox{\vskip
6pt}\hskip 3pt\vrule width 0.6pt}\hrule height 0.6pt}\kern1pt}

\begin{document}

\renewcommand{\Large}{\large}
\renewcommand{\huge}{\large}

\baselineskip = 18pt

\begin{titlepage}
\baselineskip .15in
\begin{flushright}
WU-AP/65/96
\end{flushright}

~\\

\vskip 1.5cm
\begin{center}
{\bf

\vskip 1.5cm
{\large\bf Chaos in  Static Axisymmetric Spacetimes II : non-vacuum case}

}\vskip .8in

Yasuhide {\sc Sota}$^{(a)}$, Shingo {\sc  Suzuki} $^{(b)}$and 
Kei-ichi {\sc Maeda}$^{(c)}$\\[.5em] 
{\em Department of Physics, Waseda
University, Shinjuku-ku, Tokyo 169, Japan}
\end{center}
\vfill
 \begin{abstract}
We examine the effect of local matter on the chaotic behavior
 of a relativistic test particle in non-vacuum
static axisymmetric spacetimes. 
We find that the sign of the sectional curvature in the geodesic deviation
 equation defined by the Riemann curvature does not always 
become a good tool to judge
the occurrence of chaos in the non-vacuum case.
However, we show that 
the  locally unstable region ( LU region ) defined by the Weyl
 curvature  can provide information about chaos
even in non-vacuum spacetime as well as in vacuum spacetime.
 Since the Weyl tensor affects only the shear part of the geodesic congruence,
  it works effectively to stretch some directions of geodesic congruence,
   which helps to cause the chaotic behavior of geodesics.
Actually, the orbit moving around an unstable periodic orbit (UPO) 
becomes strongly chaotic 
if it passes through an LU region, which means that the LU region can be
used as a good tool to know in which situation the chaos by
homoclinic mixing occurs around a UPO.
\end{abstract}

\vfill
\begin{center}
October, 1996
\end{center}
\vfill
(a)~~electronic mail :  694L5116@cfi.waseda.ac.jp\\
(b)~~electronic mail :  696L5186@cfi.waseda.ac.jp\\
(c)~~electronic mail :  maeda@cfi.waseda.ac.jp\\
\end{titlepage}
20

%
%
%
%
%
%
%
\section{Introduction} 
General Relativity (GR) is the most plausible theory of gravitation to explain relativistic astrophysical phenomena in our universe. It has succeeded in explaining many observational results in cosmology and
astronomy by using idealized models such as Friedmann Robertson-Walker
(FRW) spacetime or Schwarzschild spacetime.\par
However these kinds of idealized models cannot cover all of the astrophysical phenomena which might actually happen in our universe.
 Chaos may be one of these phenomena because these idealizations usually force a strong symmetry on spacetime, which 
extinguishes the possibility of complicated behavior such as chaos.
 In a Hamiltonian system chaos can occur only in the non-integrable system where  the number of integrals is less than the dimension of configuration space\cite{nonlinear}.

For example, a Bianchi-IX universe, one of the spacetimes which 
is generalized from a closed FRW universe by dropping isotropy, 
is known to show stochastic behavior \cite{BianchiIX}.
However, there has been much discussion about the definition
of chaos independent of the choice of time coordinate \cite{gauge}.

As for test particle motion in GR, the Kerr-Newman spacetime, which
 includes the Schwarzschild, Kerr and Reissner-Nordstr\"{o}m spacetimes 
as special cases, is known to be integrable. 
This can be explained by the fact that it is a Petrov type D spacetime
 and in addition to the energy $E$, angular momentum $L$ and super-hamiltonian 
$H\equiv\frac{1}{2}g^{\mu\nu}p_{\mu}p_{\nu}$, 
there is a fourth conservative quantity described by the Killing tensor \cite{Carter}. 
Thus bound orbits in this spacetime are strictly constrained to a torus in the phase space, and none of them behaves chaotically.
However, in realistic situations the spacetime around compact
 objects such as a black hole or a neutron star is not expected to have such a high degree of symmetry, because of the distortion of the gravitational
 source itself or of the effect of other gravitational sources such as gravitational waves,
 magnetic fields or other astronomical objects.
  There is a possibility that the deviations of these spacetimes from Petrov D extinguish
 the Killing tensor and cause strong chaos in the test particle motion around those compact
 objects \cite{Ge}.
 
 In fact more generic spacetimes are usually non-integrable and some
of them exhibit chaotic behavior in test particle motions \cite{Ge}-\cite{Cornish}. 
In GR,  an unstable periodic orbit (UPO) exists in axisymmetric and
static spacetime in contrast with Newtonian mechanics. This is one
of the main factors which causes chaos for geodesics in the spacetime.
For example, 
chaos is caused by periodic perturbations around  a UPO in an integrable static and axisymmetric spacetime, such as the Schwarzschild spacetime \cite{Bombelli}\cite{Moeckel}.
 Although Schwarzschild spacetime is static, spherically symmetric and integrable, the perturbations extinguish the time symmetry or axisymmetry. This type of chaos can be explained by
 the homoclinic tangle around a UPO \cite{Homo}. 
 Such chaos can also occur by non-periodic perturbations
like adding a small gravitational source to the spacetime, because it brings about the asymmetry around
 a UPO \cite{sota}\cite{Vieira}.
On the other hand, it is not trivial to determine whether or not chaos occurs  if
there is no perturbation around a UPO. 
It seems necessary to find tools to judge where and with what strength chaos occurs in these spacetimes, in order to make a quantitative analysis of chaos.

In the previous paper (paper I \cite{sota}), we paid attention to the fact that in GR, free particle motion in a spacetime is described by a geodesic, so that local instability may be determined from the sign of
the curvature tensor. 
We showed that in axisymmetric static vacuum spacetimes, geodesics
become strongly chaotic after passing through an LU region defined by the distribution of eigenvalues of the Weyl tensor. 
Recently, Vieira and Letelier showed, with their careful
analysis around UPO, that chaos predicted
by the LU region also originates from a homoclinic tangle, so that
the local instability criterion is not so strict as to predict the
chaotic property for the orbit passing through an LU region  \cite{Vieira2}.
This result suggests that in addition to the passage
through an LU region, it seems necessary for the orbit to pass by the UPO
to cause strong chaos.
However, from a physical viewpoint, the origin of the chaos around a UPO is
still unclear without information about the LU region. For example,  
why the torus suddenly begins to break around UPO in Fig.5 of \cite{sota}
 cannot be explained just by the homoclinic tangle.
Since the LU region can be utilized as an effective tool for judging the occurrence of chaos even in such a case, 
it is still important to analyze the correlation between  the LU region and
chaos, and to
give its physical meaning. 

 In the previous paper, we examined the eigenvalues of the Weyl tensor, because in vacuum spacetime
the Weyl tensor coincides with the Riemann tensor which is used to determine the local
instability in the geodesic deviation equation. 
In the non-vacuum case, however, the situation is not simple, because the Riemann tensor is no longer the same as the Weyl tensor. 
In this paper, to see the effect of matter, we first analyze the deviation of geodesics
by dividing the Riemann tensor 
into the Weyl tensor, the Ricci tensor and the scalar curvature in section 2.
This procedure makes it clear how the eigenvalues of the Riemann tensor relate to those of the Weyl tensor. 
In section 3, we will numerically show that the LU region defined by the Weyl tensor still remains a good tool to determine chaos even in spacetimes with a matter field. We will also show that even if the local matter effect is
strong enough to make the sectional curvature ${\cal K}(u,n)$ negative everywhere in
bound regions, chaos can still occur around an LU region.
This result supports the claim of Vieira and Letelier on the weakness
of the local instability determined by the sign of ${\cal K}(u,n)$.
In section 4, in order to reveal the physical origin of such chaos, we will analyze this type of chaos from another viewpoint, the shear effect of the geodesic congruence,
and show the role of the Weyl tensor on the eigenvalues of the shear matrix.
We will show that the LU region has the character of possessing two independent
stretch directions for geodesic congruence, which might explain the
contribution of the LU region to the chaos of geodesics passing
by a UPO.
 Finally, we will give our conclusions and some remarks in section 5.
%
%
%
\section{Eigenvalues of the Riemann tensor}
%
%
%
\subsection{Derivation of the eigenvalues of the Riemann tensor }
In this paper, we use the same notation as in paper I \cite{sota}. 
In GR, the motion of a free particle is described by a geodesic and
its deviation $n^\mu$ is given by the equation
\beq
 \frac{D^2n^\mu}{D\tau^{2}}= -R^\mu_{\: \nu\rho\sigma}
u^\nu n^\rho  u^\sigma,
\label{deviat}
\eeq
where $u^\mu$ is the 4-velocity and $n^\mu$ is orthogonal to $u^\mu$, i.e., $u^\mu u_\mu =-1$, and $u^\mu n_\mu =0$.
Equation (\ref{deviat}) is also reduced by the sectional curvature ${\cal K}(u,n) (\equiv -R(u,n,u,n)/\|n\|^2)$ of the plane
 spanned by $u$ and $n$ to,
\barr
\frac{d^2}{d\tau^{2}}\| n\| ={\cal K}(u,n)\| n\| 
+{1 \over 2\| n\|} \left\| n \times {Dn \over D\tau}
\right\|^2, 
\label{deviat2}
\earr
where
we use the notation $\|V\| \equiv (|V_\mu
V^\mu|)^{1/2}$ for a 4-vector
$V^\mu$. From (\ref{deviat2}), if ${\cal K}(u,n)$ is positive, a geodesic
 becomes locally unstable since the second term in the right hand
 side of (\ref{deviat2}) is always positive.
As described in paper I,
 the sectional curvature takes critical values in
the direction of the eigenvectors of the Riemann tensor.
Here we will show how these eigenvalues are expressed in a
static axisymmetric spacetime.

 In general, the Riemann tensor $R_{\mu\nu\rho\sigma}$ can be decomposed by using the Weyl tensor $C_{\mu\nu\rho\sigma}$, the Ricci tensor $R_{\mu\nu}$ and the Ricci scalar $R$ into the form
\beq
R_{\mu\nu\rho\sigma}=C_{\mu\nu\rho\sigma}+g_{\mu[\rho}R_{\sigma]\nu}-g_{\nu[\rho}R_{\sigma]\mu}
-\frac{1}{3}R g_{\mu[\rho}g_{\sigma]\nu}\label{riemann1}.
\eeq
In bivector formalism, the Riemann and Weyl tensors are regarded as $6 \times 6$ matrices
which are decomposed as follows:
\barr
{\cal R}=\left( \begin{array}{cc} \tilde{{\cal E}}&\tilde{{\cal H}}
\\-\tilde{{\cal H}^T}&\tilde{{\cal F}}\end{array}\right),  ~~~
{\cal C}=\left( \begin{array}{cc} {\cal E}&{\cal H}\\-{\cal
H}&{\cal E}\end{array}\right)
\label{matrix},
\earr
where ${\cal E}$, $\tilde{{\cal E}}$, ${\cal H}$, $\tilde{{\cal H}}$, ${\cal F}$ and $\tilde{{\cal F}}$ are $3\times3$ matrices \cite{MTW}.
In the vacuum case the parts composed of the Ricci tensor and Ricci scalar vanish from Einstein's equation, and 
the Riemann tensor coincides with the Weyl tensor.
However, in the non-vacuum case, we do not have such an advantage.
We have to deal with both ${\cal R}$ and ${\cal C}$, separately.
If the spacetime is static, the matrices
${\cal R}$ and ${\cal C}$ are expressed  as follows:
\barr
{\cal R}=\left( \begin{array}{cc} \tilde{{\cal E}}&0\\0&\tilde{{\cal F}}\end{array}\right),  ~~~
{\cal C}=\left( \begin{array}{cc} {\cal E}&0\\0&{\cal
E}\end{array}\right)
\label{32.15}
\earr
In this case, we can always diagonalize the matrices ${\cal R}$ and ${\cal C}$
 by using an appropriate orthonormal real tetrad basis
 $\{ e_{(0)}, e_{(1)}, e_{(2)}, e_{(3)}\}$
 since the matrices ${\cal E}$, $\tilde{{\cal E}}$ and  $\tilde{{\cal F}}$ in 
${\cal R}$ and ${\cal C}$ are all symmetric. 
We denote the six eigenvalues of  ${\cal R}$ as follows
\barr
\kappa^R_{0i}&=&-R(e_{(0)}, e_{(i)}, e_{(0)}, e_{(i)}) ~~~(i<j, ~i, j =1 \sim 3)\non\\
\kappa^R_{jk}&=&R(e_{(j)}, e_{(k)}, e_{(j)}, e_{(k)}) = \kappa^R_{ji}
~~~(j<k, ~j, k =1 \sim 3).
\earr
Then, the sectional curvature ${\cal K}(u,n)$ is
expressed by $\kappa^R_{\mu\nu}$ as follows.
\barr
{\cal K}(u,n)&\equiv &-R(u, n, u, n)
= -u^{(\alpha)}n^{(\beta)}u^{(\gamma)}n^{(\epsilon)}
R(e_{(\alpha)},e_{(\beta)},e_{(\gamma)},e_{(\epsilon)})\non\\ 
&=&\sum_{\mu>\nu}S_{\mu\nu} (u^{(\mu)}n^{(\nu)}-u^{(\nu)}n^{(\mu)})^2
\kappa^R_{\mu\nu},
\label{kapp1}
\earr
where $S_{\mu\nu}$ is $1$ if $\nu$ is equal to
zero, and otherwise it is $-1$.
From (\ref{kapp1}), a positive value of $\kappa^R_{0i}$ contributes to the 
local instability ( ${\cal K}(u,n)>0$), while a positive value of $\kappa^R_{ij}$ contributes to the local stability ( ${\cal K}(u,n)<0$).

For static and axisymmetric spacetimes, the metric is written as
\beq
ds^2=-e^{2U}dt^2+e^{-2U}[\rho^2 d\phi^2+
e^{2k}(d\rho^2+dz^2)],
 \label{metric}
\eeq
where $t$ and $\phi$ are the coordinates related to 
two Killing vectors $\partial/\partial t$ and
$\partial/\partial \phi$ and both $U$ and $k$ are  functions depending 
only on $\rho$ and $z$
{\renewcommand{\thefootnote}{\fnsymbol{footnote}}
\footnote[3]{ We use units of $G=c=1$, but we  explicitly
write $G$ or $c$ when it may help our discussion}.
In these coordinates, $e_{(0)}$ and $e_{(3)}$ coincide with the normalized 
Killing vectors $e_{\hat{0}}$ and
$e_{\hat{3}}$, respectively.
Then $\kappa^R_{03}$ and $\kappa^R_{12}$  always become ${\cal R}^{3}_{\: 3}$ and
${\cal R}^{6}_{\: 6}$, respectively and
the other eigenvalues are 
\barr
\kappa^R_{01} & = & \frac{1}{2} \left[ ({\cal R}^2_{\: 2}+{\cal R}^1_{\: 1}) +
\sqrt{({\cal R}^2_{\: 2}-{\cal R}^1_{\:1})^2+4({\cal R}^2_{\: 1})^2}\right]
,\non\\
\kappa^R_{02} & = & \frac{1}{2}\left[ ({\cal R}^2_{\: 2}+{\cal R}^1_{\: 1}) - 
\sqrt{({\cal R}^2_{\: 2}-{\cal R}^1_{\:1})^2+4({\cal R}^2_{\: 1})^2}\right],
\label{koyuu2}
\earr
where the matrix ${\cal R}^A_{\: B}$ is the Riemann tensor $R$ in bivector formalism and the components $A$ and $B$ of ${\cal R}^A_{\: B}$ run
$1 \sim 6$  which denote the tetrad components  ($\hat{t}\hat{\rho}$), ($\hat{t}\hat{z}$), ($\hat{t}\hat{\phi}$), ($\hat{z}\hat{\phi}$), ($\hat{\phi}\hat{\rho}$), and  ($\hat{\rho}\hat{z}$),
respectively. 
$\kappa^R_{23}$ and $\kappa^R_{31}$ can be derived by changing the indices $1$ and $2$ in (\ref{koyuu2}) to $4$ and $5$, respectively.
We can also get the eigenvalues of ${\cal C}$ using the same tetrad basis
as follows,
\barr
\kappa^c_{0i}&=&-C(e_{(0)}, e_{(i)}, e_{(0)}, e_{(i)}) \non\\
\kappa^c_{jk}&=&C(e_{(j)}, e_{(k)}, e_{(j)}, e_{(k)})=\kappa^c_{kj},
~~~(j<k, ~j, k =1 \sim 3).
\earr
Here we have $\kappa^c_{0i}=\kappa^c_{jk}$.
For the metric (\ref{metric}),
these eigenvalues, $\kappa^c_{\mu \nu}$ are derived by changing the Riemann tensor ${\cal R}$ in (\ref{koyuu2}) into the Weyl tensor ${\cal C}$ as
\barr
\kappa^c_{01}=\kappa^c_{23} & = & \frac{1}{2} \left[ {\cal C}^1_{\:
1} + {\cal C}^2_{\: 2} + \sqrt{({\cal C}^1_{\: 1}-{\cal
C}^2_{\:2})^2+4({\cal C}^1_{\: 2})^2}\right] ,\non\\
\kappa^c_{02}=\kappa^c_{31} & = & \frac{1}{2}\left[{\cal C}^1_{\:
1} + {\cal C}^2_{\: 2} -  \sqrt{({\cal C}^1_{\: 1}-{\cal
C}^2_{\:2})^2+4({\cal C}^1_{\: 2})^2}\right]
\label{koyuu3}
\earr
and $\kappa^c_{03}=\kappa^c_{12}={\cal C}^3_{\: 3}$.
In \cite{sota}, we called a region where
$\kappa^c_{03}<0$, $\kappa^c_{02}>0$ and $\kappa^c_{01}>0$ a locally unstable (LU) region, since $\kappa^c_{01}>0$ and $\kappa^c_{02}>0$
means that the sectional curvatures spanned by $e_{(0)}$ and one of the two eigenvectors
orthogonal to two Killing directions, ${\cal K} (e_{(0)},e_{(1)})$ and ${\cal K} (e_{(0)},e_{(2)})$ are positive.
 In fact, we showed in \cite{sota} that this region strongly affects the chaotic behavior of free particle motion. 

In the non-vacuum case, however, the eigenvalues of the Riemann tensor
are not the same as those of the Weyl tensor and no longer degenerate. We cannot divide the spacetime by the eigenvalues of Riemann tensor as simply as in the vacuum case, because we have six independent eigenvalues.

 In order to judge the chaos from curvature information, we have to know the relationship among these eigenvalues
and clarify how the effect of local matter changes the value of 
$\kappa^c_{\mu\nu}$ from that of $\kappa^R_{\mu\nu}$.
By using the relationship between the Riemann tensor and the Weyl tensor
(\ref{riemann1}) and Einstein's equation $G_{\mu\nu}=8\pi T_{\mu\nu}$, we get the relationship between $\kappa^c_{\mu\nu}$ and
$\kappa^R_{\mu\nu}$ (see Appendix A for the detail).
\barr
\kappa^R_{0i}&=&\kappa^c_{0i}
-\frac{4\pi}{3}\{ \rho_0-p_i+2(p_j+p_k) \} \non\\
\kappa^R_{jk}&=&\kappa^c_{jk}
+\frac{4\pi}{3}\{ 5(\rho_0-p_i)-2(p_j+p_k) \}
\label{riemann7},
\earr
where $\rho_0 \equiv T(e_{(0)}, e_{(0)})$ and $p_i \equiv T(e_{(i)}, e_{(i)})$.

 From (\ref{kapp1}), both positive values of $\kappa^R_{0i}$ and negative
 values of $\kappa^R_{jk}$ contribute to make ${\cal K}(u,n)$ positive.
In general we cannot give any concrete inequality between
$\kappa^c_{\mu \nu}$ and $\kappa^R_{\mu \nu}$ by specifying energy conditions.
 
However, 
 we can find that the electromagnetic field always weakens the local instability determined by the Weyl curvature.
That is, in this case $T$ vanishes, so that (\ref{riemann6}) in Appendix A becomes 
\barr
\kappa^R_{0i}&=&\kappa^c_{0i}-4\pi\{ \rho_0-p_i \}\non\\
\kappa^R_{jk}&=&\kappa^c_{jk}+4\pi\{ \rho_0-p_i \}
 ~~~(i, j, k =1 \sim 3)
\label{riemann8}
\earr
and $\rho_0-p_i$ is always positive,  because of the dominant energy condition \cite{Wald}.  Hence the eigenvalues $\kappa^R_{0i}$ are always
 smaller than $\kappa^c_{0i}$, while $\kappa^R_{ij}$ is always bigger than $\kappa^c_{ij}$.

 If the matter effect is not strong enough to change 
the sign of $\kappa^c_{02}$ or $\kappa^c_{03}$ in the LU region, 
the region where both $\kappa^R_{02}$ and $\kappa^R_{03}$ are positive,
 which we call the RLU region, remains inside the LU region.
Then we may use the RLU region for a criterion of chaos.
However, if the RLU region does not exist inside the LU region,
it seems to be difficult to explain chaos around the LU region by the
positivity of ${\cal K} (u,n)$. In fact in the vacuum case, the LU region was
characterized
as the region where ${\cal K} (e_{(0)},e_{(1)})>0$ and ${\cal K} (e_{(0)},e_{(2)})>0$ in \cite{sota}. However
it is nothing but the RLU region, if a matter field exists.

In the next section, we will numerically examine the relationship between the chaotic behavior of a geodesic and its passage through an LU or RLU region by using some analytically given spacetimes.
%
%
%
%
%
%
\section{Numerical analysis}
Here we examine the relationship between chaos and LU or RLU regions
by numerical calculations for some spacetimes
with matter fields.
First, we reanalyze the
Majumdar-Papapetrou(MP) spacetimes as examples of
spacetimes where both LU and RLU regions appear. 
Then, we study
the Ernst universe as an example of a spacetime where
no RLU region exists inside the LU region.
Finally, we treat
the spacetime with a scalar field as an example  of the spacetime with
a matter field which
does not satisfy any energy condition.

\subsection{Chaos around N-maximally charged black holes}
First, we examine the 
MP solution. This describes
  the system of several extreme Reissner-Nordstr\"{o}m (RN) black
holes that are balanced by the attractive gravitational and the
repulsive electric forces. 
 In this spacetime, $N$ black holes can be located at random in a three 
dimensional hypersurface.
The metric is \cite{Chandra},
\barr
ds^2&=&-V^{-2}dt^2+V^2(dx^2+dy^2+dz^2) \label{Maju1}\\
V&=&1+\sum_{i=1}^N \frac{M_i}{r_i},
\label{Maju2}
\earr
where $M_i$ is the mass of the $i$th black hole and $r_i\equiv[(x-x_i)^2+(y-y_i)^2+(z-z_i)^2]^{1/2}$
 is the Euclidean coordinate distance from the location of the $i$th black hole, $(x_i, y_i, z_i)$,
to a point $(x, y, z)$.
If $N = 1$, it is just the extreme
 RN solution. Since the RN solution belongs to Petrov type D, two eigenvalues are degenerate everywhere and neither LU nor RLU regions appear. 
This is consistent with the fact that stationary axisymmetric Petrov D spacetimes are integrable and chaos does not occur there. 

The case of $N>1$ has been examined by several authors \cite{Conto}\cite{Cornish}.
They indicated that these spacetimes cause chaotic type motions characterized by fractal boundaries of initial conditions for both timelike and null geodesics on the meridian plane. \par
Here we examine the case with the $N$ black holes located on the $z$ axis, since we are interested only in the axisymmetric case. In this case, the spacetime is described by using the two functions $U$ and $k$
 in the metric (\ref{metric})  with
\barr
U&=&-\ln (1-U_C),\non\\
k&=&0,
\label{curz}
\earr
where $U_C$ corresponds to that of $N$-Curzon spacetime \cite{Curzon} and is \beq
U_C=-G\sum_{i=1}^{N}\frac{M_{i}}{r_{i}}
\label{ncul}.
\eeq
First, we study the case $N=2$ with non-zero angular momentum of a test particle, $L$. 
In this case, we can easily find an LU region as in 2-Curzon or 2-ZV spacetime \cite{sota}, since each of two gravitational sources on the $z$ axis attracts the geodesics toward itself (Fig.\ref{2black.fig}(a)). Then if the energy of test particle is appropriately large, the bound orbit can be overlapped with LU region and strong chaos occurs beyond some critical energy (Fig.\ref{2black.fig}(b)). Here we can also find a RLU region inside an LU region, since the effect of the electric field produced by the electric charge of each black hole is not so strong as
to extinguish it.

Next, we consider the case of $N=3$, with one of the black holes located at the origin  and the others at the same distance on both sides of the
$z$-axis. In this case, we can also see an RLU region inside an LU region.
When the mass of the central black hole is much more than that of
the others, the LU region does not overlap the bound region. 
We have not seen
 any chaotic behavior there, even if we increase
  the energy up to the value of the UPO.
   However, as the mass of the central black hole reduces to that of the others, 
   the bound region are more and more overlapped with the LU region.
   Here we analyzed the case that the energy of geodesic is a little higher
   than the energy of UPO, $E_{UPO}$, where the classically admitted   region does not become
   a bound region, but has the tiny throat connecting it with a black hole.
  In this case, we can check whether or not the chaos determined by
 an LU region always has the origin of homoclinic mixing as Vieira and Letelier claimed, since the orbit passing by the throat falls into a black hole.
   In this case when the orbit passes through the LU region, 
the torus in Poincar\'e map is completely broken(Fig.\ref{3black.fig}),
although the orbit falls
into black hole with a finite time interval.
These results certainly suggest that passing through the LU region 
(and RLU region) contributes to the occurrence of chaos (though it must be defined with
finite time interval)
as for geodesic motion
in a multi-black hole system, regardless of homoclinic tangle. 

\subsection{Chaos around the Ernst universe}
Next, we reexamine the Ernst solution which includes a magnetic field.
Axisymmetric stationary spacetimes with magnetic field  $B_0$ along the $z$ axis can be derived from the corresponding vacuum spacetimes
\cite{Ernst}.
In particular, when we start with the metric form (\ref{metric}) for the static vacuum case, we find the function $U_B$ and $k_B$ of the spacetime with magnetic field to be :
\barr
U_B&=&U\ln{(1+\frac{1}{4}B_0^2 \rho^2 e^{-2U})}\non\\
k_B&=&k+2\ln{(1+\frac{1}{4}B_0^2 \rho^2 e^{-2U})},
\label{trans4}
\earr
where $B_0$ is the magnitude of the magnetic field along the $z$ axis.

Several years ago, Karas and Vokrouhlicky found that both neutral and charged particles behave chaotically in one of 
the Ernst solutions, i.e, around a Schwarzschild black hole immersed in a magnetic field \cite{Karas}.
 These particles show chaotic behavior
more and more strongly as the energy increases
 and approaches the critical value, $E_{UPO}$, beyond which the particles fall into the black hole.
This suggests that the chaos occurs through the homoclinic mixing
around a UPO.
However the origin of such a homoclinic mixing is still unclear, 
at least in the respect that the metric has a reflection symmetry on the equatorial plane to retain a homoclinic orbit 
around the UPO.
Here we examine whether it is possible to explain the chaos 
by passing through an LU region.
We reanalyze the case which they examined,
that is, the case that $U$ and $k$ correspond to the functions of
Schwarzschild spacetime and $B_0=0.15~/M$.
 In Fig.\ref{ernst.fig} we   plot  the LU region and bound
region of a neutral test particle with $(E^2,~L)=(29.12~\mu ^2,~25.0~\mu M)$, where $\mu$ is
 the rest mass of a test particle. 
From this figure, we can see that the magnetic field $B_0$ plays a role in causing an LU region
 around a single black hole, since no LU region appears
for the case of $B_0=0$. 
Certainly, it is possible to explain the strong chaos shown in \cite{Karas}
by the passage through an LU region,
since the orbit moves inside the bound region almost ergodically
and passes through the LU region regardless of the initial conditions. 
On the other hand, we cannot see an RLU region anywhere inside an LU region, because the matter effect is strong enough to change the sign of  $\kappa^R_{01}$ or $\kappa^R_{02}$ into negative through equation (\ref{riemann8}).
This result makes it difficult to explain chaos by local instability determined by geodesic deviation equation,
since the local instability in the $\rho$-$z$ plane is determined by $\kappa^R_{01}$ and $\kappa^R_{02}$, rather than $\kappa^c_{01}$ and $\kappa^c_{02}$. 

So here, in order to see the effect of the absence of an RLU region,
we consider the eigenvalues of the following matrices,
\barr
\hat{{\cal R}}^{\mu}_{\: \rho}&=&R^\mu_{\: \nu\sigma\rho}
u^\nu   u^\sigma 
\non \\
\hat{{\cal C}}^{\mu}_{\: \rho}&=&C^\mu_{\: \nu\sigma\rho}
u^\nu   u^\sigma.
\label{weylmatrix}
\earr
The sectional curvature, ${\cal K}(u,n)$ for any pair of $(u,n)$ is 
described by the linear combination of the non-zero eigenvalues $\alpha_{\hat{i}}
,\: (i=1 , \ldots , 3)$
of matrix $\hat{{\cal R}}$ as follows,
\beq
{\cal K} (u,n)=\sum_{\hat{i} =1}^3 \alpha_{\hat{i}}\: (n^{\hat{i}})^2.
\label{realsec}
\eeq
(See Appendix B.)
In order to fix the 4-velocity $u$ in (\ref{weylmatrix}),
 we use the parameter $\Theta_{\ast}$ defined as,
\barr
u^{(\hat{1})}&=&v_{\ast}\cos{\Theta_{\ast}} \non\\
u^{(\hat{2})}&=&v_{\ast}\sin{\Theta_{\ast}},
\earr
where $v_{\ast}$ is the meridian velocity which we defined in \cite{sota} as follows. 
\beq
v_{\ast}^2(\mbox{\boldmath $x$}, E, L)=
\frac{(E^2-V^2_{\rm eff}(\mbox{\boldmath $x$}, L))}
{2\|\partial/\partial t \|^{2}},
\label{vhosi}
\eeq
where 
$\mbox{\boldmath $x$}=(\rho, z)$ 
and $V^2_{\rm eff}$ is
 the effective potential
 for the particle with the angular momentum
$L$.
The rest components, $u^{(\hat{0})}$ and $u^{(\hat{3})}$ are determined by $E$ and $L$,
respectively.

We numerically examined the distribution of eigenvalues $\alpha_{\hat{i}}$ 
in the bound region and
found that all of the eigenvalues $\alpha_{\hat{i}}$ become negative everywhere inside the bound region for any
pair of $(v_{\ast}, \Theta_{\ast})$  with $(E^2,~L)=(29.12~\mu^2,~25.0~\mu M)$. This means that chaos cannot be explained by the curvature term in (\ref{deviat}) or (\ref{deviat2}), since ${\cal K} (u,n)$ becomes negative
for any pair of $(u,n)$ from (\ref{realsec}).
This result seems inconsistent with the result
that strong chaos occurs for the same value of $E$ and $L$,
which leads us to new analysis by a shear effect of the geodesic
congruence on chaos rather than that of the geodesic deviation, as we will see in section 4. 
\subsection{The spacetime with scalar field}\eqnum{0}
Finally, we examine a spacetime with a scalar field, as an example
of a spacetime with a matter field which satisfies neither the
strong nor dominant energy condition. In this case, an RLU region need not be included
in the LU region, since the matter field may increase the local instability.

Here we make use of the solution in which $N$-scalar charged
ZV-type singularities are located on the $z$ axis \cite{Azuma}.
We examine the case of
two scalar charged singularities balanced against
one another on the $z$ axis, by the gravitational force and the repulsive force induced by the scalar charge.
This balance condition is attained by choosing each scalar charge so as to satisfy the relationship 
$ M_i=\Sigma_i (i=1,2)$, where $M_i$ is the mass of the $i$th singularity and $\Sigma_i$ is its scalar
charge. This condition is similar to that of the MP solution, where extreme black holes are balanced together.
In this case, the metric is described by setting $k=0$
in the corresponding 2-ZV solution, 
where $M_i=m_i\delta$ ((3.10) in \cite{sota}) and this spacetime becomes asymptotically flat.
As we showed in section 2, the RLU region is not always included in the LU region in general, in contrast with the case with an electromagnetic field. In fact, the RLU region can even be seen outside the LU region (Fig.\ref{scalar.fig}).
However, the local matter effect is not strong enough to
separate the RLU region from the LU region completely, so that
chaos occurs for a particle passing through both of these regions.
Hence, in this case
we cannot clearly
determine through which region  chaos occurs, that is, through the LU or the RLU region. 
%
%
%
%
\section{The shear effect of the LU region on geodesic congruence}\eqnum{0} 
In section 3.2, we showed that no RLU region
appears inside an LU region and ${\cal K} (u,n)$ becomes negative everywhere inside the bound region in spite of the occurrence of 
strong chaos.
This failure of the ${\cal K} (u,n)$ criterion may come from the fact that
the positivity of  ${\cal K} (u,n)$ is a sufficient condition 
for the local instability of geodesic 
but may not be necessary condition, because of the second term in the right side of (\ref{deviat2}).
This second term  could induce an instability of $\| n\|$
 against the negative contribution of the ${\cal K} (u,n)$ term. 
 In this  case, we could not explain chaos simply by a
local instability determined from the sign of ${\cal K} (u,n)$. 
On the other hand, the LU region still works well to judge the occurrence
of chaos even in the Ernst case,
 since any chaotic bound trajectory passes through an LU region.

In order to see the effect of an LU region even in this case, we 
examine the deviation of nearby geodesics by
expansion $\theta$ and shear $\sigma_{\mu\nu}$ of the geodesic congruence.
It is well known that $\theta$ and $\sigma_{\mu\nu}$ satisfy the following
equations \cite{Wald},
\barr
\frac{d\theta}{d\tau}&=&-\frac{1}{3}\theta^2-\sigma_{\mu\nu}\sigma^{\mu\nu}
-R_{\mu\nu}u^{\mu}u^{\nu}, 
\label{thetayo}\\
\frac{D\sigma_{\mu\nu}}{D\tau}&=&-\frac{2}{3}\theta \sigma_{\mu\nu}
-\sigma_{\mu\rho}\sigma^{\rho}_{\: \nu}+\frac{1}{3} h_{\mu\nu}\sigma_{\rho\epsilon}\sigma^{\rho\epsilon}
+\hat{{\cal C}}_{\mu\nu}+\frac{1}{2} R^{(TF)}_{\mu\nu},
\label{sigmayo}
\earr
where $R^{(TF)}_{\mu\nu}$ is the trace-free part of $R_{\mu\nu}$ and defined by
\beq
R^{(TF)}_{\mu\nu}=h_{\mu\rho}h_{\nu\sigma}R^{\rho\sigma}-\frac{1}{3}h_{\mu\nu}h_{\rho\sigma}R^{\rho\sigma},
\label{tildeR}
\eeq
where $h_{\mu\nu}\equiv g_{\mu\nu}+u_{\mu} u_{\nu}$ is the metric
of 3 space perpendicular to $u_{\mu}$.
Here we assume that the twist term $\omega_{\mu\nu}$  vanishes and 
examine the time evolution of a geodesic congruence whose
initial value of $\theta, \sigma_{\mu\nu} =0$.
From (\ref{thetayo}), if the Ricci tensor $R_{\mu\nu}$ satisfies
the condition 
\beq
R_{\mu\nu}u^{\mu}u^{\nu} \geq 0
\label{energy1}
\eeq
for an arbitrary 4-velocity $u$, $\theta$ negatively diverges, which is followed by the creation of a conjugate point of the
geodesic.  From Einstein's equation $G_{\mu\nu}=8\pi T_{\mu\nu}$, the condition (\ref{energy1}) is attained, if and only if the energy-stress tensor $T_{\mu\nu}$ satisfies the following strong energy condition
(see \cite{Wald}).
\beq
T_{\mu\nu}u^{\mu}u^{\nu} \geq -\frac{1}{2}T.
\label{energy2}
\eeq
As long as the condition (\ref{energy2}) is satisfied, geodesic congruence will converge everywhere. 
On the other hand, the above energy condition tells us nothing about the
shear of the geodesic congruence. It is possible for the
geodesic to be stretched exponentially by the shear term $\sigma$
even in a spacetime with a matter field satisfying the strong energy condition (\ref{energy2}).
This stretching property of shear seems to be
 an important cause of chaos, as we see in the famous baker's transformation \cite{Gulick}. Since the Weyl tensor directly affects
the time evolution of shear in (\ref{sigmayo}), it is expected to play an important role in chaos through the stretching of nearby geodesics in combination with their bending at the edge of the bound region.

Especially for spacetimes with an electromagnetic field, since
the time derivative of expansion $\theta$ is negative
in (\ref{thetayo}), the volume of geodesic congruence always converges  because of the strong
energy condition (\ref{energy2}). So shear may be one of the important factors that causes
local instability for geodesics in spacetimes with an electromagnetic field.
In the equation of shear (\ref{sigmayo}), only the last two terms on
the right side,
\beq
\Omega^{\mu}_{\: \nu}=\hat{{\cal C}}^{\mu}_{\: \nu}
+\frac{1}{2} R^{\mu\: (TF)}_{\: \nu}
\label{Omega}
\eeq
are independent of the initial values of $n$ and $Dn/D\tau$. Hence
$\Omega^{\mu}_{\: \nu}$ is important when examining the effect of shear on chaos, 
since chaotic behavior of a trajectory is generally independent of 
the initial choice of $n$ around a given geodesic.
Since each part of (\ref{sigmayo}) is trace-free, the sum of all eigenvalues of $\Omega^{\mu}_{\: \nu}$ 
vanishes. 
Hence one of the eigenvalues of $\Omega^{\mu}_{\: \nu}$ is always positive and diverges
as the geodesic approaches the gravitational source singularity, since this stretch comes from 
the tidal force by the gravitational source. 

For the asymptotically flat spacetime with a single gravitational source singularity, the largest eigenvalues of $\Omega^{\mu}_{\: \nu}$,
 $\zeta_{1}$, always becomes positive and decreases monotonically to zero as $r \rightarrow \infty$
  (Fig.\ref{one-bla.fig}).
The direction of its corresponding eigenvector projected on the $\rho$-$z$ plane is almost parallel
to that of $u$ (Fig.\ref{eigenvec.fig}). (Note that those eigendirections are orthogonal to $u$ in the 4-dimensional spacetime because of conditions I and II in Appendix B.)
 The rest of the eigenvalues
always become negative and monotonically increase to zero as $r \rightarrow \infty$.
These properties for the eigenvalues of $\Omega^{\mu}_{\: \nu}$ also hold for those of
the Weyl tensor, since the tidal force in the direction of the  gravitational source is characterized by the positive eigenvalue of the Weyl tensor. 
On the other hand, in multi-black holes spacetimes or Ernst spacetime, one
of the negative eigenvalues of the Weyl tensor becomes positive inside the LU region. (In fact, the sign change of this eigenvalue  defines the LU region.)
 Since $\hat{{\cal C}}^{\mu}_{\: \nu}$ is defined by the Weyl 
tensor (\ref{weylmatrix}),
one of the eigenvalues of $\hat{{\cal C}}^{\mu}_{\: \nu}$ also becomes positive in the LU region. 
In the non-vacuum case,  the local matter part
 $\frac{1}{2} R^{\mu\: (TF)}_{\: \nu}$ should also be considered in $\Omega^{\mu}_{\: \nu}$.
In the Ernst case, the local matter effect $\frac{1}{2} R^{\mu\: (TF)}_{\: \nu}$ in (\ref{Omega})
 is not strong enough to change the sign of the two
positive eigenvalues, $\zeta_1$ and $\zeta_2$ of $\Omega^{\mu}_{\: \nu}$ in the LU region back to negative 
 (Fig.\ref{ernstomega.fig}). Moreover, the eigendirections of $\Omega^{\mu}_{\: \nu}$ are almost the same as those of $\hat{{\cal C}}^{\mu}_{\: \nu}$ and the Weyl tensor.

In order to see the effect of its two positive eigenvalues on shear,
we compare the eigendirections of $\Omega^{\mu}_{\: \nu}$ with those
of $\sigma^{\mu}_{\: \nu}$. As we can see in Fig.\ref{eigencomp.fig},
these eigendirections are almost the same, which means that the
two positive eigenvalues of the Weyl tensor have the effect of
stretching the geodesic congruence in two independent directions through
the shear effect. 
As we can see in Fig.\ref{eigenvec.fig}, the second eigendirection whose eigenvalue
 changes from negative to positive is not parallel to $u$ on the $\rho$-$z$ plane,
  in contrast with the first eigendirection of $\zeta_{1}$.
It is suggested that the second positive eigenvalue of the Weyl tensor contributes to stretching the congruence in the direction independent of  $u$ and causes chaos 
for bound orbits passing through the LU region. 

To substantiate this speculation,  we examine  the magnitude of the positive eigenvalue $\zeta_{2}$ of 
$\Omega^{\mu}_{\: \nu}$ inside a bound region 
and compare it with the value of the Lyapunov exponent $\lambda$ of chaotic motion
inside this region.
From Fig.\ref{peak.fig}(a), we can see that for a given value of $L$, the peak of $\zeta_{2}$ becomes higher and higher, as $E$ 
gets larger. This is consistent with the
result in our previous paper \cite{sota} that the Lyapunov exponent $\lambda$, that is, the
strength of chaos, is proportional to $v_{\ast}$, since the effect on
shear becomes larger as $v_{\ast}$ increases. 
For concrete values of $\lambda$, in the Ernst case with $(E^2,~L)=(29.12~\mu^2,~25.0~\mu M)$,
we find the numerical value of $\lambda\sim 3.15\times10^{-2}/M$, regardless of initial conditions, which
is the same order of magnitude as the peak value of $\zeta_{2}$.

As for the $\Theta_{\ast}$ dependence of $\zeta_{2}$,
its peak becomes larger and larger as the geodesic crosses
 the equatorial plane orthogonally, although the change
is not so large as to change the order of the value (Fig.\ref{peak.fig}(b)). 
We also find the same results in the relationship between shear and the LU region
for multi-black hole systems.
Thus these results substantiate our claim that
the shear effect in the eigendirection corresponding to $\zeta_{2}$ is correlated with the chaotic motion of the geodesic. 
%
%
%
%
\section{Concluding Remarks}\eqnum{0} 
In this paper, we have examined the LU region criterion for chaos in the non-vacuum case, by analyzing test particle motion in several axisymmetric static spacetimes 
as well as in the vacuum case.

In multi-black hole systems, the RLU region was always seen inside the LU region, while in a spacetime with 2 scalar-charged singularities, the RLU region was seen even outside the LU region.
In both cases, we found that passing through an LU or RLU region
gives us good information about the occurrence of chaos,
although it is indeterminate
which region is more fundamental for the chaos because of 
the slight separation of these regions.

In an Ernst universe, however, the RLU region cannot be seen, because one
of the two positive eigenvalues 
of the Weyl tensor is made negative by the strong local matter effect.
In this case
the local instability of geodesics is not explained by the curvature term in (\ref{deviat}) or (\ref{deviat2}), since all of the eigenvalues of $\hat{{\cal R}}^{\mu}_{\: \nu}$ become negative,
which means that
${\cal K} (u,n)$ becomes negative for any direction of $(u,n)$ at
any point in bound region. 
Then we examined the character of the LU region
 to explain such a chaotic behavior  by dividing
the matrix $\hat{{\cal R}}^{\mu}_{\: \nu}$ into the shear part,
 $\Omega^{\mu}_{\: \nu}$ and the expansion part $R_{\mu\nu}u^{\mu}u^{\nu}$.
We find that the eigenvalues of $\Omega^{\mu}_{\: \nu}$ are strongly affected by those of the Weyl tensor, so that 
in addition to the positive eigenvalue determined by the tidal force
of gravitation, the second eigenvalue of $\Omega^{\mu}_{\: \nu}$ also becomes positive around the LU region.
It follows from this property that two eigenvalues of the shear matrix $\sigma^{\mu}_{\: \nu}$
also become positive in almost the same directions as the eigen
directions of $\Omega^{\mu}_{\: \nu}$, as we numerically showed in sec.4. 
Thus the LU region is characterized as the region in which the geodesic congruence 
is stretched in the direction independent of $u$ in addition to the direction of $u$.
We can speculate that this additional stretch helps to cause strong chaos for geodesics passing through the LU region, especially around UPO. 
In fact we found that the Lyapunov exponent
of those chaotic motions can be estimated by the peak of the additional second type eigenvalue.
These results suggest that the Weyl tensor  plays a much more important role in determining the chaos of geodesics than the sign of sectional curvature ${\cal K}(u,n)$.

 Strictly speaking, the chaos for the orbit passing 
through the LU region cannot completely be determined
 by the shear effect.
For example, we cannot explain,  just by the LU region criterion, why the chaos occurs only for the orbit whose energy
is almost near to $E_{UPO}$ and
why non-chaotic orbit coexists with chaotic orbit even if both of them
pass through LU region for the fixed value of $E$ and $L$
(see, for example, Fig.2 in \cite{Vieira2}.)
As Vieira and Letelier showed in \cite{Vieira2}, 
the chaos also seems correlated with the homoclinic tangle around the
UPO. It might be because the stretching property of the LU region
helps homoclinic mixing and makes the occurrence of chaos 
much easier around UPO. Hence, by our criterion we can just give
a good tool to find a chaotic orbit.

Another interesting point we wish to stress is as follows.
As we showed in sec.3.1, such a chaotic behavior can also be seen
  for the orbit with the energy beyond $E_{UPO}$,  which cannot 
  be explained by the homoclinic tangle. In a strict sense, we cannot define the chaos 
in such an unbounded case, since the orbits remain within the bound region only in
the finite time interval. However, this result suggests that
the LU region or
the eigenvalues of the Weyl tensor could have some correlation with
the unbounded chaotic type phenomena such as chaotic scattering
 or a fractal basin \cite{Cornish}\cite{scattering}.
 Further analysis alomg this line will be left for the future work.

Here we restricted ourselves to the static case, where there is no rotation 
effect of spacetime involved in the geodesic equation.
However, it will be interesting to extend our analysis to rotating 
spacetimes
not only for the free particle motion but also for spinning particles \cite{Shingo}, because
 the spin-orbit or spin-spin interaction may provide new physical ingredient. 
All of these speculations lead us to future works which may reveal some important roles for chaos in realistic  relativistic astronomical phenomena in our universe. 

\vskip 1cm

\noindent
-- Acknowledgments --
We would like to thank R. Easther for his critical reading of our
paper. This work was supported partially by the Grant-in-Aid for Scientific 
Research Fund of the Ministry of Education, Science and Culture  
(No. 06302021 and No. 06640412) and by the Waseda University Grant 
for Special Research Projects. 

\newpage

\appendix

\section{Appendix A}

Starting from the Einstein's equation $G_{\mu\nu}=8\pi T_{\mu\nu}$,
 we rewrite Eq.(\ref{riemann1}) using the energy stress tensor $T_{\mu\nu}$ as follows
\beq
R_{\mu\nu\rho\sigma}=C_{\mu\nu\rho\sigma}+4\pi \{ g_{\mu[\rho}T_{\sigma]\nu}- g_{\nu[\rho}T_{\sigma]\mu}\}+\frac{8\pi}{3}Tg_{\mu[\rho}g_{\sigma]\nu}  
\label{riemann2}.
\eeq
Then, we obtain
\beq
R(u,n,u,n)=C(u,n,u,n)+4\pi \{ -{\cal T}(n,n)
+\| n\|^2{\cal T}(u,u)\}+ \frac{8\pi}{3}T\| n\|^2.
\label{riemann3}
\eeq
With the definition ${\cal K}(u,n) \equiv - R(u,n,u,n) /\| n\|^2 $ and ${\cal K}_c(u,n) \equiv - C(u,n,u,n)/ \| n\| ^2$, Eq.(\ref{riemann3}) is now
\beq
{\cal K}(u,n)={\cal K}_{c}(u,n)
+4\pi \left\{ \frac{{\cal T}(n,n)}{\| n\|^2}-{\cal T}(u, u)\right\} 
- \frac{8\pi}{3}T,
\label{riemann4}
\eeq
 where ${\cal T}(u, u)$ represents the energy density $\rho_0$ as measured by the observer whose 4-velocity is $u$ and ${\cal T}(n, n)/\| n \|^2$ represents
 the principal pressure $p_n$ in the direction of $n$\cite{Wald}. Then (\ref{riemann4}) becomes
\beq
{\cal K}(u,n)={\cal K}_{c}(u,n)-4\pi \{ \rho_0-p_n \}- \frac{8\pi}{3}T
\label{riemann5}
\eeq
Note that it is not guaranteed  that the space-space components of $T_{\mu\nu}$ are always diagonalized \cite{ Hawking}. However, it is easily shown that as long as the Riemann tensor is diagonalized, 
 its components are inevitably diagonalized
by the principal pressure $p_i$ for the tetrad basis $\{ e_{(0)}, e_{(1)}, e_{(2)}, e_{(3)}\}$. Hence by contracting two of the tetrad basis elements $\{ e_{(0)}, e_{(1)}, e_{(2)}, e_{(3)}\}$ with Eq.(\ref{riemann2}),
we find the relation
between the eigenvalues of the Weyl tensor and those of Riemann tensor as follows:
\barr
\kappa^R_{0i}&=&\kappa^c_{0i}-4\pi\{ \rho_0-p_i \}-\frac{8\pi}{3}T
\non\\
\kappa^R_{jk}&=&\kappa^c_{jk}+4\pi\{ \rho_0-p_i \}-\frac{8\pi}{3}T
 ~~~(i, j, k =1 \sim 3),
\label{riemann6}
\earr
 where $\rho_0 \equiv {\cal T}(e_{(0)}, e_{(0)})$ and $p_i \equiv {\cal T}(e_{(i)}, e_{(i)})$.
By substituting the connection $T=-\rho_0+\sum_{\scriptstyle i}p_i$, these components become (\ref{riemann7}).

\section{Appendix B}

We first pay attention to the following
properties of matrices, $\hat{{\cal R}}$ and $\hat{{\cal C}}$ in (\ref{weylmatrix}).
\begin{enumerate}
\renewcommand{\labelenumi}{\Roman{enumi}}
\item) One of the eigenvalue of matrix $\hat{{\cal R}}$ ( or $\hat{{\cal C}}$) is trivial, i.e., 
the corresponding eigenvector is the 4-velocity $u$ and its eigenvalue
is zero.
\item) Any two eigenvectors of matrix $\hat{{\cal R}}$ ( or $\hat{{\cal C}}$) are orthogonal each other. This is not so trivial, since matrix the $\hat{{\cal R}}$
 is not symmetric on all indices. 
\end{enumerate}
II can be proven by taking the property of the Riemann tensor 
and the definition of $\hat{{\cal R}}$ into account.

 From Property II, it is always possible to define an orthonormal triad basis, $E^{\mu}_{\hat{i}}\;(\hat{i}=1 , \ldots , 3)$, orthogonal to $u$
by using the normalized
eigenvectors of $\hat{{\cal R}}$ and expand by them the
normalized deviation vector $\hat{n}$ defined as $\hat{n}\equiv n/\| n\|$
like 
\beq
\hat{n}=\sum_{\hat{i}=1}^{3}n^{\hat{i}} E_{\hat{i}},
\label{expand}
\eeq
from the condition, $(u,n)=0$.

From (\ref{expand}) and the definition (\ref{kapp1}), ${\cal K}(u,n)$ can be described as 
\barr
{\cal K}(u,n)&=&-R(u,n,u,n) \non\\
&=&-\sum_{i=1}^{3}\sum_{j=1}^{3}n^{\hat{i}}n^{\hat{j}}
R(u,E_{\hat{i}},u,E_{\hat{j}}) \non\\
&=&\sum_{i=1}^{3}\sum_{j=1}^{3}n^{\hat{i}}n^{\hat{j}}
(E_{\hat{i}},\:\hat{{\cal R}} E_{\hat{j}}).
\label{appeB1}
\earr
From the condition $\hat{{\cal R}}E_{\hat{i}}=\alpha_{\hat{i}}E_{\hat{i}}$,
(\ref{appeB1}) becomes
\beq
{\cal K}(u,n)
=\sum_{i=1}^{3}\sum_{j=1}^{3}n^{\hat{i}}n^{\hat{j}}
\alpha_{\hat{j}}(E_{\hat{i}},E_{\hat{j}}).
\label{appeB2}
\eeq
From the condition $(E_{\hat{i}},E_{\hat{j}})=\delta_{ij}$,
we can get (\ref{realsec}).

\newpage

%
%

\newpage

\begin{figure}[h]
\caption[fig.1]
{(a) LU and RLU 
region (lightly shaded)  of the 2-black hole spacetime  with two equal
masses, $M$,  located at $\pm 2 GM/c^2$ on the $z$ axis
(black dots) and the bound region (dotted) of a test
particle with  the angular momentum $L=3.4~G\mu
M/c$ and energy, $E^2=0.664~(\mu c^2)^2$
corresponding to $E_{\rm UPO}$.
(b)Poincar\'e map for the orbit in the bound region with the initial
condition $p^{\rho}_{\: 0}=0.0, 
z_0=0$ and $\rho_0 = 3.0~GM/c^2$  }
\label{2black.fig}
\end{figure}

\begin{figure}[h]
\caption[fig.2]{(a),(b) classically admitted region in 3 black hole case with
the mass ratio of central black hole to each of outside black hole
(a)1:0.9, (b)1:0.95. The parameter for each region is $E^2=(a)0.5712, (b)0.72$ 
and $L=(a)5.2, (b)6.0$, respectively.
(c) Poincar\'e map for the orbit in the bound region of (a) with
the initial condition of each orbit is $p^{\rho}_{\: 0}=0.0, 
z_0=0$ and $\rho_0 =1.5,1.6,1.7,1.8,1.9~GM/c^2$.
Chaos cannot be seen in this case.
(d)  the orbit in the admitted region of (b) with
the initial condition,  $p^{\rho}_{\: 0}=0.0, 
z_0=0$ and $\rho_0 =1.5~GM/c^2$and (e) it's Poincar\'e map.
 The Poincar\'e map of the orbit passing through an LU region
certainly becomes chaotic, before it falls into the central black hole through
tiny throat.}
\label{3black.fig}
\end{figure}

\begin{figure}[h]
\caption[fig.3]{LU and bound region in Ernst case with
magnetic field $B_0=0.15~c^2/GM$ around Schwarzschild black hole.
The parameter for bound region is $L=25.0~G\mu
M/c$ and energy, $E^2=29.116816~(\mu c^2)^2$.}
\label{ernst.fig}
\end{figure}

\begin{figure}[h]
\caption[fig.4]{(a) LU and RLU region in the spacetime with 
scalar field, where two singularities with scalar charge $Q$ is
fixed on $z$ axis at $z\pm 2.0$. The bound region (dotted) with  $L=6.9~G\mu
M/c$ and $E^2=0.90913~(\mu c^2)^2$
corresponding to $E_{\rm UPO}$ is also depicted.
 RLU region is almost overlapped with LU region, but not included in it.
(b)Poincar\'e map for the orbit in the bound region with the initial
condition $p^{\rho}_{\: 0}=1.0, 
z_0=0$ and $\rho_0 = 2.8~GM/c^2$ .}
\label{scalar.fig}
\end{figure}

\begin{figure}[h]
\caption[one-bla]{$\zeta_{1}$ and $\zeta_{2}$ of RN solution
on equatorial plane. Here we used the parameter $\Theta_{\ast}=0$, $L=2.92$ and $(E/\mu c^2)^2=0.8663$ corresponding to $E_{\rm UPO}$. $\zeta_{1}$
decreases monotonically as the coordinate $\rho$ increases, while $\zeta_{2}$ increases monotonically and approaches to zero.}
\label{one-bla.fig}
\end{figure}

\begin{figure}[h]
\caption[eigenvec]{The distribution of eigenvector field on $\rho$-$z$ plane for the Ernst case, Fig.3 with $\Theta_{\ast}=0$. The eigenvector fields
corresponding to (a) $\zeta_{1}$ and (b) $\zeta_{2}$ are
depicted}
\label{eigenvec.fig}
\end{figure}

\begin{figure}[h]
\caption[ernstomega]{(a)$\zeta_{1}$ and $\zeta_{2}$ of $\hat{{\cal C}}^{\mu}_{\: \nu}$ for Ernst solution on equatorial plane. Here we used the parameter $\Theta_{\ast}$=0, $L=25.0~G\mu M/c$ and $(E/\mu c^2)^2=29.116816$ corresponding to $E_{\rm UPO}$. (b)$\zeta_{1}$ and $\zeta_{2}$ of $\Omega^{\mu}_{\: \nu}$ for the same condition as (a). Both of the cases show that
 $\zeta_{1}$ decreases monotonically, while $\zeta_{2}$
has the peak around LU region.}
\label{ernstomega.fig}
\end{figure}

\begin{figure}[h]
\caption[fig.5]{The direction of eigenvector on $\rho$-$z$ plane
for matrix $\sigma^{\mu}_{\: \nu}$ and $\Omega^{\mu}_{\: \nu}$
along the orbit with the initial condition, $p^{\rho}_{\: 0}=0.0, 
z_0=0$ and $\rho_0 =4.0~GM/c^2$ in the case of Fig.3. (a) and (b) are the two eigendirections for two positive eigenvalues
of $\sigma^{\mu}_{\: \nu}$ and $\Omega^{\mu}_{\: \nu}$, respectively. (c) and (d) are the two eigendirections for two negative
 eigenvalues
of $\sigma^{\mu}_{\: \nu}$ and $\Omega^{\mu}_{\: \nu}$, respectively. The tendency of the changes of positive
 eigendirections of $\sigma^{\mu}_{\: \nu}$ on $\rho$-$z$ plane  along the orbit
  is almost the same as those of $\Omega^{\mu}_{\: \nu}$. So is the tendency of the changes of negative
  eigendirections.
}
\label{eigencomp.fig}
\end{figure}

\begin{figure}[h]
\caption[fig.5]{$E^2$ and $\Theta_{\ast}$ dependence of
 $\zeta_{2}$ of matrix $\hat{{\cal C}}$ on equatorial plane for the same
  situation as fig.\ref{ernst.fig}.
(a)$E^2$ dependence of $\zeta_{2}$ for $\Theta_{\ast}$=0, $L=25.0~G\mu M/c$
 and $(E/\mu c^2)^2=$29.116816[(i)], 27.0[(ii)],26.0[(iii)]. We also added
  $\zeta_{2}$ of matrix $\hat{{\cal C}}$ in Schwarzschild spacetime for comparison,
   where we used the parameter $\Theta_{\ast}$=0, $L$=3.55
    and $(E/\mu c^2)^2=$0.9025 corresponding to $E_{\rm UPO}$.
(b)$\Theta_{\ast}$ dependence of $\zeta_{2}$ for $\Theta_{\ast}=0$[(i)],
 $0.0$ [(ii)],$\pi/6.0$[(iii)]$ \pi/3.0$, $L=25.0~G\mu
M/c$ and $(E/\mu c^2)^2=29.116816$.  As the $\Theta_{\ast}$ increases and
the direction become parallel to equatorial plane, the eigenvalue
becomes smaller and smaller. }
\label{peak.fig}
\end{figure}

\end{document}